\documentclass[preprint,showpacs,aps,preprintnumbers,amsmath,amssymb,
nofootinbib]{revtex4}
\usepackage{epsfig}
\begin{document}

\begin{flushright}
\end{flushright}


\newcommand{\be}{\begin{equation}}
\newcommand{\ee}{\end{equation}}
\newcommand{\bea}{\begin{eqnarray}}
\newcommand{\eea}{\end{eqnarray}}
\newcommand{\bers}{\begin{eqnarray*}}
\newcommand{\eers}{\end{eqnarray*}}
\newcommand{\nn}{\nonumber}
\newcommand\un{\cal{U}}
\def\dis{\displaystyle}
\def\ds{m_{D_s}^2}
\def\sss{\scriptscriptstyle}
\def\barp{{\raise.35ex\hbox
{${\sss (}$}}--{\raise.35ex\hbox{${\sss )}$}}}
\def\bark{{\raise.35ex\hbox
{${\sss (}$}}--{\raise.35ex\hbox{${\sss )}$}}}
\def\barpd{{\raise.35ex\hbox
{${\sss (}$}}--{\raise.35ex\hbox{${\sss )}$}}}
\def\dbarp{\hbox{$D^{*0}$\kern-1.35 em\raise1.5ex\hbox{\barpd}}}
\def\dbark{\hbox{$K^{*0}$\kern-1.35 em\raise1.5ex\hbox{\barpd}}}
\def\bdbarp{\hbox{$B_d^0$\kern-1.1 em\raise1.4ex\hbox{\barp}}}
\def\rp1{\hbox{$D^{*0}$\kern-1.55 em\raise1.5ex\hbox{\barpd}}}
\def\arp1{\hbox{$B_d$\kern-1.1 em\raise1.4ex\hbox{\barp}}}
\def\drp2{\hbox{$K^{*0}$\kern-1.6 em\raise1.5ex\hbox{\barpd}}}

\title{\large Possible signatures of unparticles in  rare annihilation type
$B$ decays }
\author{R. Mohanta$^1$  and A. K. Giri$^2$  }
\affiliation{$^1$ School of Physics, University of Hyderabad,
Hyderabad - 500 046, India\\
$^2$ Department of Physics, Punjabi University,
Patiala - 147002, India}

\begin{abstract}
We investigate the effect of unparticles in the pure annihilation
type decays $B^- \to D_s^-\phi $ and $ D_s^- K^{*0}$.
Since these decays have only annihilation contributions  their 
branching ratios are expected  to be very
small  in the standard model and the direct CP asymmetry parameters 
to be zero. We find that
due to the unparticle effect these branching ratios can be
significantly enhanced from their standard model values. Furthermore,
sizable nonzero direct CP violation could also be possible in these channels
due to the presence of intrinsic CP conserving phase in the unparticle 
propagator.
\end{abstract}

\pacs{14.80.-j, 11.30.Er, 13.25.Hw}
\maketitle

There are many discrete massive particles in reality which can exist
in a scale non-invariant theory as in the case of the standard model 
(SM) whereas, in a scale invariant theory,
in four space-time dimensions, the mass spectrum of fields is
continuous or zero. Recently, Goergi \cite{georgi} suggested that there could 
be a possibility that the scale invariant sector (termed as unparticle) 
indeed exists but might have been undetected so far as the 
intermediate particles mediating
these two sectors are believed to be very heavy. The standard model
fields might be scale invariant at high energy but 
scale invariance might have been
broken somewhere above the electroweak scale. However, here the
SM is taken to be scale non-invariant all the way up to high energy. 

Now let us consider a scenario in which there exist 
the scale invariant sector with non-trivial infrared fixed 
point at high energy, which is represented 
by the Banks-Zaks (${\cal BZ} $) fields \cite{bz}, and also 
the SM fields. Thus, the high energy theory
contains both the SM fields and the $B{\cal Z}$ fields, which
interact via the exchange of particles of mass $M_{\un}$ with the
generic form
\bers
\frac{1}{M_{\un}^k} O_{SM} O_{\cal{BZ}}\;,
\eers
where $O_{SM}$ is
the operator of mass dimension $d_{SM}$  and
 $O_{\cal{BZ}}$ is the operator of mass dimension $d_{\cal{BZ}}$
 made out of SM  and ${\cal{BZ}}$ fields, respectively.
The couplings of the ${\cal BZ}$ fields cause dimensional
transmutation and below the scale $\Lambda_{\un}$ the ${\cal BZ}$ 
operators match onto the unparticle
operators leading to the interaction between the SM and the Unparticle
sector of the type
\bers
C_{\un}\frac{\Lambda_{\un}^{d_{\cal{BZ}}-d_{\un}}}{M_{\un}^k} O_{SM}
O_{\un}\;,
 \eers
where $C_{\un}$ is a coefficient in the low energy effective theory
and $O_{\un}$ is the unparticle operator with scaling dimension
$d_{\un}$. Furthermore, $M_{\un}$ should be large enough such that
its coupling to the SM fields must be sufficiently weak, consistent
with the current experimental data.

Unparticle stuff  with scale dimension $d_{\un}$ looks like a
non-integral number $d_{\un}$ of invisible massless particles. 
Unparticle, if exists, could couple to the standard model fields 
and consequently affect the low energy dynamics. The effect of 
unparticle stuff on low energy phenomenology has been 
extensively explored in Refs.
\cite{kingman, ref2, roman, huang}.

A clean signal of the unparticle stuff can be inferred from various
analyses, e.g., the missing energy distribution in mono-photon production
via $ e^- e^+ \to \gamma \un $ at LEP2 \cite {kingman}  and direct
CP violation in the pure leptonic $B^\pm \to  l^\pm \nu_l$ modes etc.
\cite{roman, huang}. In this paper, we would like to see whether the pure
annihilation type $B$ decays could provide any interesting avenues
to visualize the signatures of unparticles. The decay modes
considered here are $B^- \to D_s^- \phi $ and $ D_s^-
K^{*0}$ which have only annihilation type contributions in the SM
and therefore expected to be very rare. Only the upper limits are known 
for these channels which are given as Br$(B^+ \to D_s^+ \phi)<1.9
\times 10^{-6}$ and Br$(B^+ \to D_s^+ \bar{K}^{*0}) < 4 \times 10^{-4}$
\cite{pdg}. Furthermore, since the decay amplitudes of these modes 
have contributions arising only from annihilation diagrams, 
the direct CP asymmetry parameters in these modes are identically zero. 
Although some new physics (NP) models like two Higgs doublet model  and 
R-parity violating
supersymmetric model can provide new contributions to these channels
and thereby open up the possibility of observing direct CP violation
\cite{rm}. But, if the strong phases in the SM and NP amplitudes turn
out to be tiny, CP violation cannot be observed even if there is
new physics contribution to the decay amplitudes. However, if the
new contributions to the decay amplitudes are due to the unparticles,
which contain the new CP conserving phase
coming from the unparticle propagators, CP
violation could be observed even for vanishing SM strong phase. In
this paper we would like to explore such a possibility.

The decay modes considered here are $B^- \to D_s^- \phi$ and $D_s^-
{K}^{*0}$ which arise in the standard model from the tree level annihilation 
diagrams. The $b$ and $\bar
u$ quarks of the initial $B$ meson annihilate to a $W^-$ boson which
subsequently hadronises to the final $D_s^-$  and $\phi (K^{*0}) $ mesons. 
The effective Hamiltonian describing such transitions is given as 
\bea {\cal
H}_{eff} &=& \frac{G_F}{\sqrt 2} V_{ub} V_{cq}^*\biggr[ 
C_1(\mu)~ \bar q_\alpha
\gamma_\mu(1-\gamma_5)c_\alpha ~\bar u_\beta \gamma^\mu
(1-\gamma_5)b_\beta \nn\\ &&~~~+
 C_2(\mu)~ \bar q_\beta
\gamma_\mu(1-\gamma_5)c_\alpha~ \bar u_\alpha \gamma^\mu
(1-\gamma_5)b_\beta \biggr]\;, 
\eea 
where $C_{1,2}$ are the Wilson
coefficients,  $\alpha$ and $ \beta $ are the color indices, $q=s(d)$
for the final state meson $\phi(K^{*0})$.
Thus one can obtain the transition amplitude for these processes
as
\be
 A(B^-(p_B) \to D_s^-(p_D) V(p_V, \epsilon))
 =\frac{G_F}{\sqrt 2} V_{ub} V_{cq}^*~ a_1~  X\;,
\ee
where $V$ denotes the final vector meson $\phi /K^{*0}$,  
$a_1=C_1+C_2/N_c$ and $X= \langle
D_s^- V|\bar q \gamma^\mu(1-\gamma_5)c~
 \bar u \gamma_\mu (1-\gamma_5)b|B^- \rangle$ is the 
matrix element of the four quark current operators between the
initial and final mesons. The
evaluation of this hadronic matrix element is not possible 
from the first principles
of QCD and one requires some additional assumptions to determine it.
Since, in this paper, we are interested to see the effect of unparticles
in the CP violation asymmetries, we will not pay much attention 
for its evaluation.
However, if one resorts to the generalized factorization approach
then one can factorize this matrix element as $X=\langle D_s V|
(\bar q c)_{V-A} (\bar u b)_{V-A}| B^- \rangle \equiv   \langle D_s V|
(\bar q c)_{V-A}|0\rangle \langle 0| (\bar u b)_{V-A}| B^- \rangle $.
The first element can be related to its corresponding
crossed channel as
$\langle D_s(p_D) V(p_V, \epsilon)|(\bar q c)_{V-A} |0 \rangle \equiv
\langle V(p_V, \epsilon)|(\bar q c)_{V-A} |D_s(-p_D) \rangle $\;.
Using Lorentz invariance, these matrix elements can be 
represented in terms of form factors and decay constants, as
defined in \cite{bsw}. Thus one can obtain the transition amplitude as
\bea
A(B^-(p_B) \to D_s^-(p_D)~ V(p_V, \epsilon)) &=& -\frac{G_{F}}{\sqrt 2} V_{ub} 
V_{cq}^*~a_1~
f_B ~2 m_V ~(\epsilon^* \cdot p_D)~  A_0(p_B^2)\;.
\eea
The corresponding branching ratio is given as
\bea
{\rm Br}(B^- \to D_s^- V)= \frac{p_c^3}{8 \pi m_V^2} |A( B^- \to D_s V)/
(\epsilon^* \cdot p_D)|^2 \;,
\eea
where $p_c$ is the c.o.m. momentum of the emitted particles in the
$B$ rest frame.

Now we would like to see how unparticle stuff will affect the 
transition  amplitudes. Here, we assume that the charged  unparticles 
mediate the interaction between the initial and final mesons. The 
possible existence of charged unparticles and their consequences 
have been recently discussed in Refs. \cite{roman, huang}. 
It is well known that, depending on the nature of the 
original ${\cal BZ}$ operator  $O_{\cal BZ}$ and the
transmutation, the resulting unparticle may have different Lorentz
structure. In our analysis, we consider only the scalar  and 
vector type unparticles. The coupling of
these unparticles to quarks is given as \be
\frac{c_S^{q'q}}{\Lambda_{\un}^{d_{\un}}}\bar q'
\gamma_\mu(1-\gamma_5) q~
\partial^\mu O_{\un}+\frac{c_V^{q'q}}{\Lambda_{\un}^{d_{\un}-1}}\bar q'
\gamma_\mu(1-\gamma_5) q~ O_{\un}^\mu+h.c.\;, \label{cv}
\ee where
$O_{\un}$ and $O_{\un}^\mu$ denote the scalar and vector unparticle
fields and $c_{S,V}^{q'q}$ are the dimensionless coefficients which 
in general depend on different flavors and are assumed to be real. For 
the charged unparticle
exchange, $q'$ in Eq (\ref{cv}) belongs to the up quark
sector $(u,c,t)$ and $q$ to the down type quarks ($d,s,b)$.
 
The propagator for the scalar unparticle field is given as
\cite{georgi, kingman} \be \int d^4 x e^{i P \cdot x}\langle 0 | TO_{\un}(x)
O_{\un}(0)|0 \rangle = i \frac{A_{d_{\un}}}{2 \sin d_{\un} \pi}
\frac{1} {(P^2+i \epsilon)^{2-d_{\un}}}e^{-i\phi_{\un}}\;, \ee where
\be A_{d_{\un}}= \frac{16 \pi^{5/2}}{(2 \pi)^{2 d_{\un}}}
\frac{\Gamma(d_{\un}+1/2)}{\Gamma(d_{\un}-1)\Gamma(2d_{\un})}\;,
 ~~~~{\rm and} ~~~\phi_{\un}=(d_{\un}-2)\pi \;. \label{eq5}
\ee
Similarly, the propagator for the vector unparticle is given by \be
\int d^4 x e^{i P \cdot x}\langle 0 | TO_{\un}^\mu(x) O_{\un}^\nu
(0)|0 \rangle = i \frac{A_{d_{\un}}}{2 \sin d_{\un} \pi}
\frac{-g^{\mu \nu} +P^\mu P^\nu/P^2} {(P^2+i
\epsilon)^{2-d_{\un}}}e^{-i \phi_{\un}}\;. \label{eq6}
\ee

The interesting thing in these propagators is the presence of
the new CP conserving phase $\phi_{\un}$, which leads to a  spectacular
interference pattern and hence can exhibit nonzero
direct CP violating effects in many rare decay channels. Here it 
should be noted that the vector unparticle will not contribute 
to the decay channels considered here. This is because, the contraction
of $P_\mu$ arising from axial coupling i.e., $\langle 0 | A_\mu |B^- \rangle $
 with the $(-g^{\mu \nu} +P^\mu P^\nu/P^2)$ term in the 
 vector unparticle  propagator 
gives vanishing contribution, while the vecor coupling  $\langle 0 |V_\mu
|B^- \rangle $ is identically zero. Therefore, only scalar type unparticles 
will contribute to these rare decay modes.
Thus, the contributions arising from
scalar unparticle exchange is given as
\be
 A^{\un}(B^- \to D_s^-~ V)
 = \frac{\lambda}{\Lambda_{\un}^{2 d_{\un}}}\frac{A_{d_{\un}}}{
2 \sin (d_{\un} \pi)}~(m_B^2)^{d_{\un}-1} e^{-i \phi_{\un}}~  X\;,
\ee
where $\lambda=(c_S^{ub} c_S^{cq})$ and $X$ is the hadronic matrix
element. Now, including the unparticle contributions, one can write 
the  transition amplitude as 
\bea
A^T(B^-(p_B) \to D_s^-(p_D)~ V(p_V, \epsilon)) = A^{SM}(1+r~ e^{i(\gamma
-\phi_{\un})})\;,
\eea  
where $\gamma $ is the weak phase of SM amplitude i.e., we have used 
$V_{ub}=|V_{ub}| e^{-i \gamma}$. Furthermore, $ r$ denotes 
the magnitude of the ratio of 
unparticle to SM amplitude and is given as
\be
r= \frac{\lambda}{\Lambda_{\un}^{2 d_{\un}}}\frac{A_{d_{\un}}}{
\sqrt 2 \sin (d_{\un} \pi)}~\frac{ (m_B^2)^{ d_{\un}-1}}{G_F~|V_{ub}
V_{cq}^*|~a_1}\;.
\ee
Thus, we obtain the CP averaged branching ratio $\langle {\rm Br}
\rangle \equiv [{\rm Br}(B^- \to D_s^- V)+{\rm Br}(B^+ \to D_s^+ V)]/2$, 
including the unparticle 
contributions, as
\be
\langle {\rm Br} \rangle
= {\rm Br}^{\rm SM} (1+r^2 +2 r \cos \gamma 
\cos \phi_{\un})\;,
\ee
where $ {\rm Br}^{\rm SM} $ is the
SM branching ratio. It should be noted that since the direct CP 
violation in these modes is zero in the SM, the SM CP averaged branching ratio
is same as  ${\rm Br}(B^- \to D_s^- V)$. The direct CP 
violation parameter, which is related to  the difference of the partial 
decay rates,  given as
\bea 
A_{CP} &= & \frac{ \Gamma(B^- \to D_s^-~  V) - \Gamma
(B^+ \to D_s^+ \overline{V})}{ \Gamma(B^- \to D_s^- V) + \Gamma
(B^+ \to D_s^+ \overline{V})}\nn\\
 &=& \frac{ 2 r \sin \gamma \sin \phi_{\un}}
{1+r^2 +2 r \cos \gamma 
\cos \phi_{\un}}\;.
\eea
We now proceed to evaluate the SM branching ratios using the QCD 
factorization approach. Although the QCD factorization 
predictions for annihilation contributions suffer from end point 
divergences, and therefore have large uncertainties, but one can still 
have an estimate on the order of the branching ratios. In this approach,
the annihilation amplitudes are parameterized as \cite{beneke}
\be
A(B^- \to D_s^- V)= i \frac{G_F}{\sqrt 2}V_{ub}V_{cq}^* f_B f_{D_s} f_V~
b_2(D_s, V)\;,
\ee
where $b_2=C_F C_2 A_1^i/N_c^2$, with $C_F=4/3$ as the color factor and
$N_c$ is the number of colors. $A_1^i$ is given as
\be
A_1^i \approx 6 \pi \alpha_s \biggr[ 3 \left (X_A-4+\frac{\pi^2}{3} \right )
+r_{\chi}^{D_s} r_{\chi}^V(X_A^2-2 X_A) \biggr]\;,
\ee
where
\be
r_{\chi}^{D_s}= \frac{2 m_{D_s}^2}{m_b(m_c+m_s)}\;,~~~~~~ 
r_{\chi}^{V}= \frac{2 m_{V}}{m_b}\frac{f_V^\perp}{f_V}\;.
\ee
 $X_A$ denotes the end point divergences which can be parameterized as
\be
X_A=(1+\rho_A e^{i \phi_A}) \ln\left (\frac{m_B}{\Lambda_h}\right )\;,
~~~{\rm with} ~~~\Lambda_h=0.5~{\rm GeV}\;.
\ee
For the numerical evaluation, we use the particle masses, lifetime
of $B^-$ meson and the 
CKM elements $V_{ub}=(3.96 \pm 0.09) \times 10^{-3}$, 
$V_{cs}=0.97296 \pm 0.00024$, $V_{cd}=0.2271 \pm 0.0010$ from \cite{pdg}.  
The decay constants (in GeV) as  $f_{D_s}=0.294$ \cite{pdg},
$f_\phi=0.237$ \cite{lu}, $f_B=0.2$,$f_{K^*}=0.218$, $f_V^\perp=0.175$
\cite{beneke}, the current
quark masses (in GeV) as $m_b=4.2$, $m_c$=1.3, $m_s$=0.08 \cite{beneke}. 
We use 
$C_2=-0.295 $ \cite{beneke1} 
evaluated at $m_b/2$ scale at next-to-leading order,
$\rho_A=1$ and $\phi_A=-20^\circ $ \cite{beneke} for the annihilation 
parameters.
Thus we obtain the branching ratios as
\bea
&&{\rm Br}^{\rm SM}(B^- \to D_s^- \phi) \approx  (4.2 \pm 0.2)
 \times 10^{-7}\;,\nn\\
 &&{\rm Br}^{\rm SM}(B^- \to D_s^- K^{*0}) \approx  (1.2 \pm 0.1) 
\times 10^{-8}\;,
\eea
which are well below the present upper limits and are 
also consistent with the results of 
PQCD calculation Br$(B^+ \to D_s^+ \phi)=
3.0 \times 10^{-7}$ \cite{lu} and Br$(B^+ \to D_s^+ K^{*0})=
(1.8 \pm 0.3) \times 10^{-7}$ \cite{lu1}. 
  
Now we proceed to see the effect of unparticle stuff on the branching ratios 
and CP violation parameters. As seen from Eq.(9), the unparticle contribution
contains three unknown parameters, namely, the dimension of the unparticle
fields $d_{\un}$, the energy scale $\Lambda_{\un}$ and the coupling $\lambda$. 
Therefore, to see the effect of unparticles we fix the
energy scale as $\Lambda_{\un}=1 $ TeV. Now using 
the CKM angle $\gamma=70^\circ$ we  show in figures 
1 and 2, the variations of the
branching ratios and CP violation parameters 
with $d_{\un}$  for different values of $\lambda$.
From the figures we can see that when the scale dimension $d_{\un}$
is less than 1.4, the branching ratios are very sensitive to it and above
$d_{\un}= 1.4$ the unparticle contributions are negligible and the 
branching ratio curves converge to the corresponding SM values. Furthermore,
one can also obtain significant direct CP asymmetries due to the
unparticle contributions. As seen from the figures, when the
coupling $\lambda \approx {\cal O}(1)$, maximum CP violation
i.e., up to $50\%~ (70 \%) $ is possible  for the $B^- \to D_s^- \phi~
(B^- \to D_s^- K^{*0})$ channels. These CP violation parameters 
are also quite sensitive to the scale dimension $d_{\un}$ and above 
$d_{\un} \approx 1.6 $ they approach to zero, which is the corresponding SM 
value. 
 
\begin{figure}[htb]
   \centerline{\epsfysize 2.0 truein \epsfbox{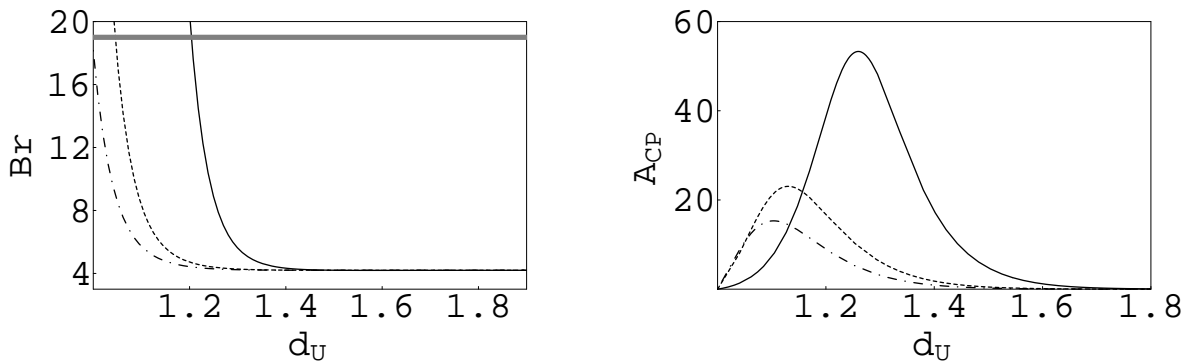}}
\caption{ CP averaged 
  branching ratio $\langle {\rm Br} \rangle$ (in units of $10^{-7}) $ and 
CP violation parameter
(in $\%$) for
the decay mode $B^- \to D_s^- \phi $, where the solid, dashed
and dot-dased lines correspond to $\lambda$=1, 0.1 and 0.05
respectively. The horizontal thick line in the branching ratio plot 
represents the experimental upper limit.}
  \end{figure}

\begin{figure}[htb]
   \centerline{\epsfysize 2.0 truein \epsfbox{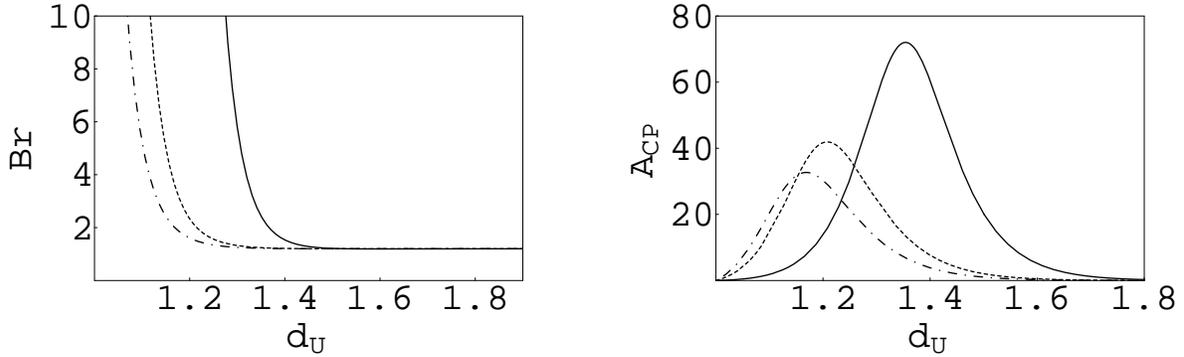}}
 \caption{ CP averaged branching ratio $\langle {\rm Br} \rangle$
 (in units of $10^{-8})$ and CP 
violation parameter (in $\%$) for
the decay mode $B^- \to D_s^- K^{*0} $. The solid, dashed
and dot-dased lines correspond to $\lambda$=1, 0.1 and 0.05
respectively.}

  \end{figure}

Unparticle physics associated with a hidden scale
invariant sector at a higher energy scale with a nontrivial
fixed point  has received significant attention in a short span of time.
Many interesting unparticle specific 
signatures have been
pointed out. In this work, we have explored another situation
where we can expect clean unparticle signatures. We have 
studied the phenomenology of unparticle physics in the rare
$B$ decay channels $B^- \to D_s^- \phi~
(B^- \to D_s^- K^{*0})$. Since in the SM these decay modes have only 
annihilation contribution so the branching ratios are
expected to be  quite small. These values
can be significantly enhanced from their SM values if unparticle effect
is taken into account. The direct CP asymmetries for these modes are 
identically zero in the SM and therefore are very much suitable ones 
to look for new 
physics. Because of the presence of intrinsic 
CP conserving phase in the unparticle propagator
nonzero and sizable direct CP violation can be observed in these modes.
To conclude, enhancement in the branching ratios in the rare
annihilation type $B$ decays $B^- \to D_s^- \phi~ (B^- \to D_s^- K^{*0})$ and
nonzero direct CP violation will be clean signals of new physics and
unparticle physics will be a strong contender for the same.

\acknowledgments The work of RM was partly supported by Department
of Science and Technology, Government of India, through grant No.
SR/S2/HEP-04/2005. AG would like to thank Council of Scientific and
Industrial Research, Government of India, for financial support.

\end{document}